\documentclass[twocolumn,prl,aps]{revtex4}
\usepackage{graphicx}
\usepackage{textcomp}
\usepackage{color}
\begin{document}

\newcommand{\CSi}{C$_{\rm 60}$}
\newcommand{\EF}{$\rm E_F$}
\newcommand{\Er}{$\rm E_r$}
\newcommand{\Vs}{$\rm V_s$}
\newcommand{\dIdV}{$\rm dI/d V_s$}
\newcommand{\Pdec}{$\rm P_{dec}$}
\newcommand{\PdecM}{$\rm \overline{P}_{dec}$}
\newcommand{\Idec}{$\rm I_{dec}$}
\newcommand{\Tm}{$\rm T_{m}$}
\newcommand{\Vdec}{$\rm V_{dec}$}
\newlength{\FigureWidth}
\setlength{\FigureWidth}{0.82 \columnwidth}

\title[Electron heating and cooling of a single \CSi\ molecule]
{Resonant heating and substrate-mediated cooling of a single \CSi\
molecule in a tunnel junction}

\author{Gunnar Schulze, Katharina J. Franke and Jose Ignacio Pascual}

\affiliation{Institut f\"{u}r Experimentalphysik, Freie
Universit\"{a}t Berlin, Arnimallee 14, 14195 Berlin, Germany}

\date{\today}

\begin{abstract}
We study the influence of different metallic substrates on the
electron induced heating and heat dissipation of single \CSi\
molecules in the junction of a low temperature scanning tunneling
microscope. The electron current passing through the molecule
produces a large amount of heat due to electron-phonon coupling,
eventually leading to thermal decomposition of the fullerene cage.
The power for decomposition varies with electron energy and reflects
the resonance structure participating in the transport. The average
value for thermal decomposition of \CSi\ on Cu(110) amounts to 21
$\mu$W, while it is much lower on Pb(111) (2.9 $\mu$W) and on
Au(111) (1.0 $\mu$W). We ascribe this difference to the amount of
charge transfer into \CSi\ upon adsorption on the different
surfaces, which facilitates molecular vibron quenching by
electron-hole pair creation.
\end{abstract}

{ \color{magenta} \pacs{73.40.-c, 73.61.Wp, 33.15.Mt} }

\maketitle

\section{Introduction}

A single molecule junction that is exposed to the flow of an
electron current will experience an increase of temperature due to
the heat generated by   conduction electrons. Recent theoretical
studies predicted that Joule heating in molecular junctions can be
large enough to affect the reliability of molecular devices
\cite{Galperin07}. The temperature of a molecular junction is
difficult to estimate. Recent experimental approaches have found
agreement in that a single molecule device can easily reach several
hundreds of degrees  under normal  operation (flow of microwatts
when powered with 1 Volt)
\cite{HuangNL06,Berndt07a,HuangNN07,Schulze08}. For most molecules,
such temperature reaches the limit of thermal degradation.
Therefore, a deeper knowledge of the microscopic mechanisms of heat
generation and dissipation on the nanoscale is required in order to
improve the performance of a molecular device.

The temperature at the junction is a consequence of an equilibrium
between heating and dissipation of heat away from the molecule. Heat
generation is essentially caused by scattering of electrons with
molecular vibrations. Heat dissipation is expected to follow several
mechanisms involved in the coupling of the hot molecule with the
substrate acting as an external cold bath. According to this, the
temperature reached by a single molecule during electron transport
should depend on the lead's material. Furthermore, one would expect
that both the phonon bandwidth and the electronic density of states
around the Fermi energy  should govern \textit{a priori} the
mechanisms of heat dissipation (cooling) and, hence, the molecular
temperature.

In this paper we analyze the effect of the lead's material on the
heat dissipated during electron transport through a single molecule.
We use a low temperature scanning tunneling microscope (STM) to
inject an electron current through a single fullerene adsorbed onto
clean metal surfaces of different nature. Rather than measuring the
temperature reached by the single molecule, we follow the approach
described in our previous work \cite{Schulze08}. We look for certain
current and sample bias (\Vs)  values at which the molecule under
the STM tip undergoes an irreversible thermal degradation. The
applied power required to reach the decomposition limit (\Pdec)
varies slightly with the electron energy (e\Vs) in a sequence that
resembles the energy alignment of the fullerene resonances for every
metal substrate, confirming a resonant mechanism of electron heating
\cite{PecchiaPRB07}. However, we also find that the mean values of
\Pdec\ (\PdecM) depend more strongly on the substrate onto which the
fullerenes are adsorbed. The way that \PdecM\ scales for different
metals cannot be explained in terms of intrinsic properties of each
material. Instead, the observed trend seems to be a result of the
mechanism of interaction between metal and molecule. In particular,
we find a correlation between  the behavior \PdecM\ and the amount
of charge transfer into molecular unoccupied states, as resolved
using local spectroscopy measurements of \CSi\ molecules on the
different substrates. Hence, our results suggest that charge
transfer processes at the metal-molecule contact provide an
effective pathway to dissipate heat from a hot molecule through the
creation of e-h pairs.

\section{Adsorption of \CSi\ on Cu(110), Pb(111) and Au(111)}
\label{Adsorption}

\begin{figure}[h]
  \begin{flushright}
    \includegraphics[width = 8cm]{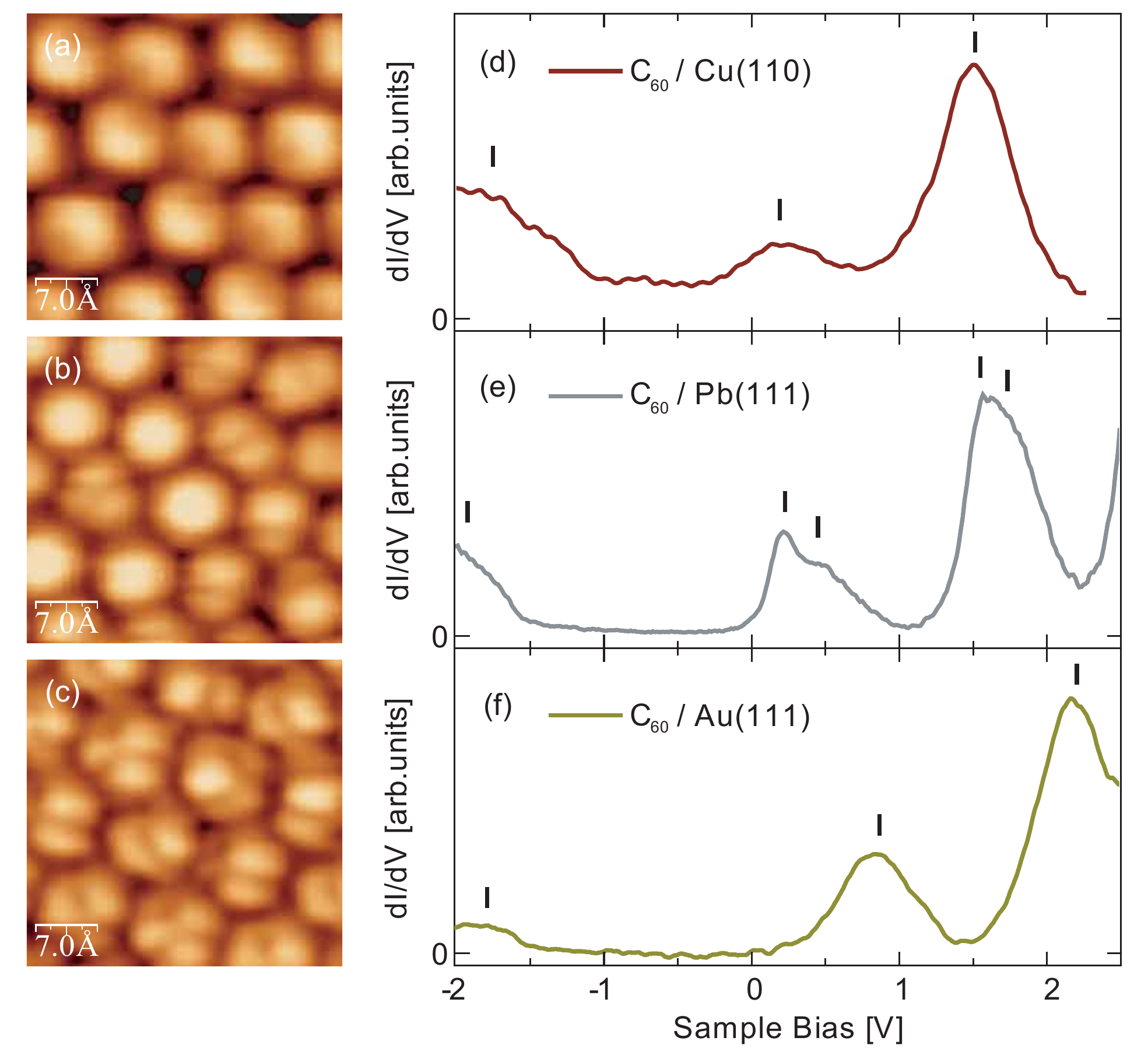}
  \end{flushright}
  \caption{(a--c): STM images of \CSi\ islands on Cu(110)
  ($I_t = 1.0\ nA$, $V_s = 2.25\ V$), Pb(111) ($I_t = 2.0\ nA$, $V_s = 0.5\ V$),
  and Au(111) ($I_t = 0.6\ nA$, $V_s = 0.4\ V$). (d-e):
  Corresponding tunnel conductivity spectra (\dIdV) of \CSi\ on the different surfaces. Bars mark
  the fitted positions of HOMO, LUMO and LUMO+1. Two peaks were fitted
  to the LUMO and LUMO+1 levels on Pb(111). The spectra were measured
  by positioning the STM tip on top of a single molecule and ramping
  \Vs while keeping the tip-molecule distance constant
  (feedback-loop open). \dIdV\ data were obtained by using a lock-in amplifier with a rms modulation amplitude $\rm V_{ac}$.
  (Cu: $R_{junct} = 1.1\  G\Omega$, $V_{ac} = 20\ mV$, Pb: $R_{junct} = 0.7 \
  G\Omega$, $V_{ac} = 5\ mV$, Au: $R_{junct} = 0.3\ G\Omega$, $V_{ac} = 30\ mV$.)}
  \label{ImagedIdV}
\end{figure}

The experiments were performed in a custom-made ultra-high vacuum
STM, which is operated at a temperature of 5 K.  We chose Cu(110),
Pb(111) and Au(111) single crystals as substrates  because of their
different properties regarding chemical reactivity, as well as
phonon band widths and densities of states at the Fermi energy
(\EF). Cleaning of the metal substrates was performed by standard
sputtering-annealing cycles under ultra-high vacuum, ensuring an
atomically clean and flat metal surface. Indentations of the STM tip
into the substrate while applying a tip-sample bias were used to
clean the tips in regular intervals. The tips are therefore believed
to be composed of the same  material as the substrate. A
sub-monolayer coverage of \CSi\ molecules was deposited from a
Knudsen cell on the metal surfaces at room temperature. On Cu(110)
the system was further annealed to 470 K to ensure that the
molecules self-assemble in ordered domains and populate a thermally
stable adsorption state. The structure and electronic configuration
of the molecular layers were characterized using STM and scanning
tunneling spectroscopy (STS) measurements (Fig. \ref{ImagedIdV}).

On Cu(110), \CSi\ forms ordered islands with a pseudo-hexagonal
structure (Fig. \ref{ImagedIdV} (a)), in which  the fullerenes adopt
a well-defined adsorption configuration with a pentagon-hexagon C-C
bond pointing upwards \cite{Schulze08,FaselPRB99}. An STS spectrum
on these molecules (Fig. \ref{ImagedIdV} (d)) resolves  a clear
spectroscopic fingerprint characterized by a sharp resonance at
$\sim$1.5 eV above the Fermi level and associated to the alignment
of the LUMO+1 resonance (LUMO: lowest unoccupied molecular orbital).
The LUMO resonance appears as a broader peak centered at $\sim$0.2
eV and is partially occupied.

Fullerene adsorption on Pb(111) at room temperature results in
highly ordered hexagonal islands (Fig. \ref{ImagedIdV} (b)). Here,
scanning tunneling spectroscopy reveals the LUMO and LUMO+1 derived
resonances as in Cu(110) but more pronounced and with a
characteristic split structure due to the breaking of degeneracy
upon adsorption \cite{note} (Fig. \ref{ImagedIdV} (e)). The
energetic alignment of the LUMO close to \EF, with  a small tail
crossing it, indicates a small amount of charge transfer into \CSi.

\CSi\ islands on Au(111) evidence a similar hexagonal lattice as on
Pb(111) (Fig. \ref{ImagedIdV} (c)), but with large variety regarding
molecular orientations \cite{Berndt07c}, as one can determine from
intramolecular structure resembling the lobed shape of the LUMO
resonance \cite{PascualJCP02}. Both LUMO and LUMO+1 resonances are
resolved in STS spectra as pronounced peaks, independently of the
molecular orientation \cite{RogeroJCP02}. The  LUMO peak appears
located far from \EF, at $\sim$ 0.8 eV, thus indicating that charge
transfer from the substrate is very small \cite{CrommiePRB04}.

\section{\CSi\ decomposition on metal surfaces}
\label{Decomposition}

In our experiment, we approach the STM tip a distance Z towards a
single \CSi\ molecule while holding the sample bias \Vs\ constant.
During approach we record the current flowing through the molecular
junction ($\rm I(Z)$). The tunnel regime is revealed by an
exponential increase of $\rm I(Z)$ with diminishing gap distance. At
a certain approach distance the $\rm I(Z)$ curves deviate smoothly
from the exponential dependence, indicating the onset of  a
tip-molecule contact \cite{Berndt07a}. The conductivity at this
point is a small fraction of $\rm G_0$ ($G_0 = 77.5\ \mu S$). For
small positive sample bias  the molecule remains intact upon contact
formation and even further indentations of several {\AA}ngstroms,
leading to a stable junction with the molecule contacted by the STM
tip on the one side, and the metal surface, on the other
\cite{Berndt07a,Schulze08}.  In this case, the integrity of the
indented molecule after tip contact can be verified by its
appearance in the STM images and, especially, by its electronic
fingerprint in \dIdV\ spectra.

\begin{figure}[h]
  \begin{center}
    \includegraphics[width = 8cm]{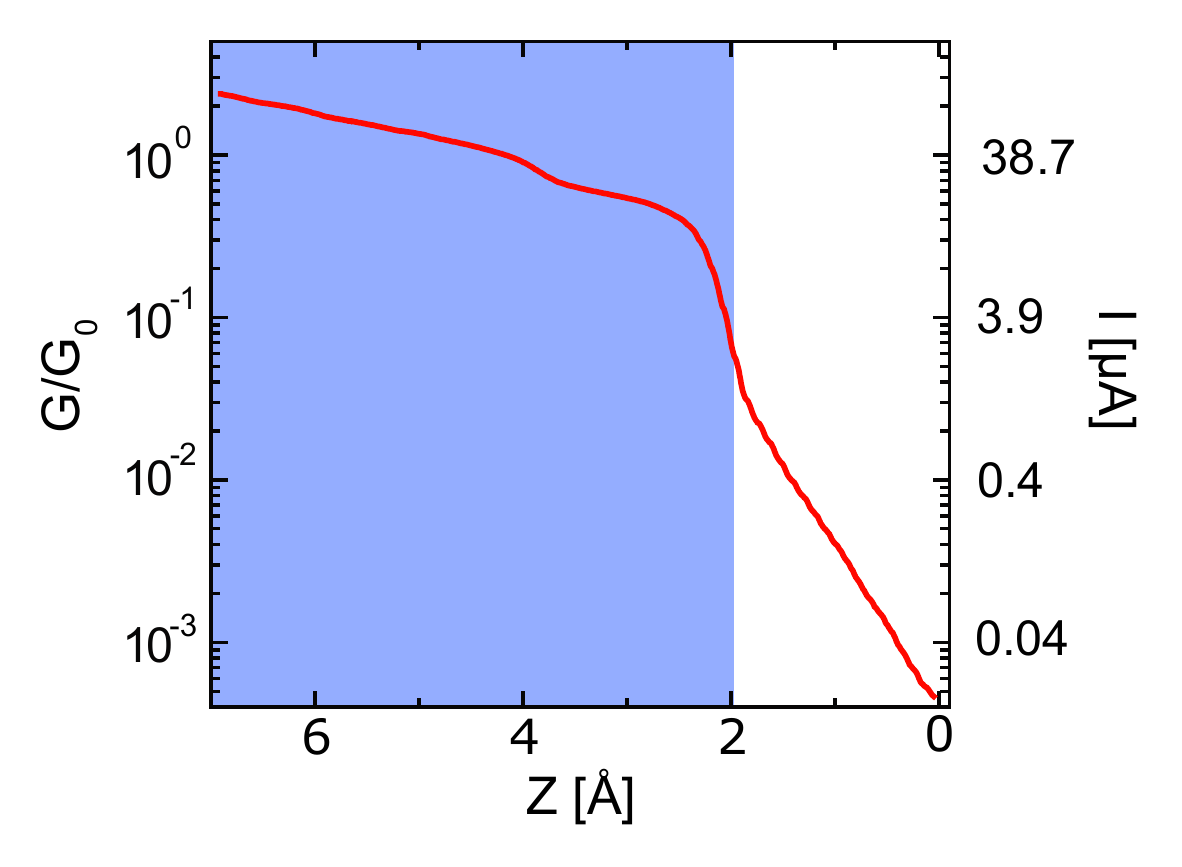}
  \end{center}
  \caption{Conductance and current vs. Z-distance plot I(Z) on top of a
  \CSi\ molecule on Cu(110) ($V_s = 0.5\ V$). The blue shaded area indicates the contact
  regime. }
  \label{Indentation}
\end{figure}

\begin{figure}[h]
  \begin{flushright}
    \includegraphics[width = 8cm]{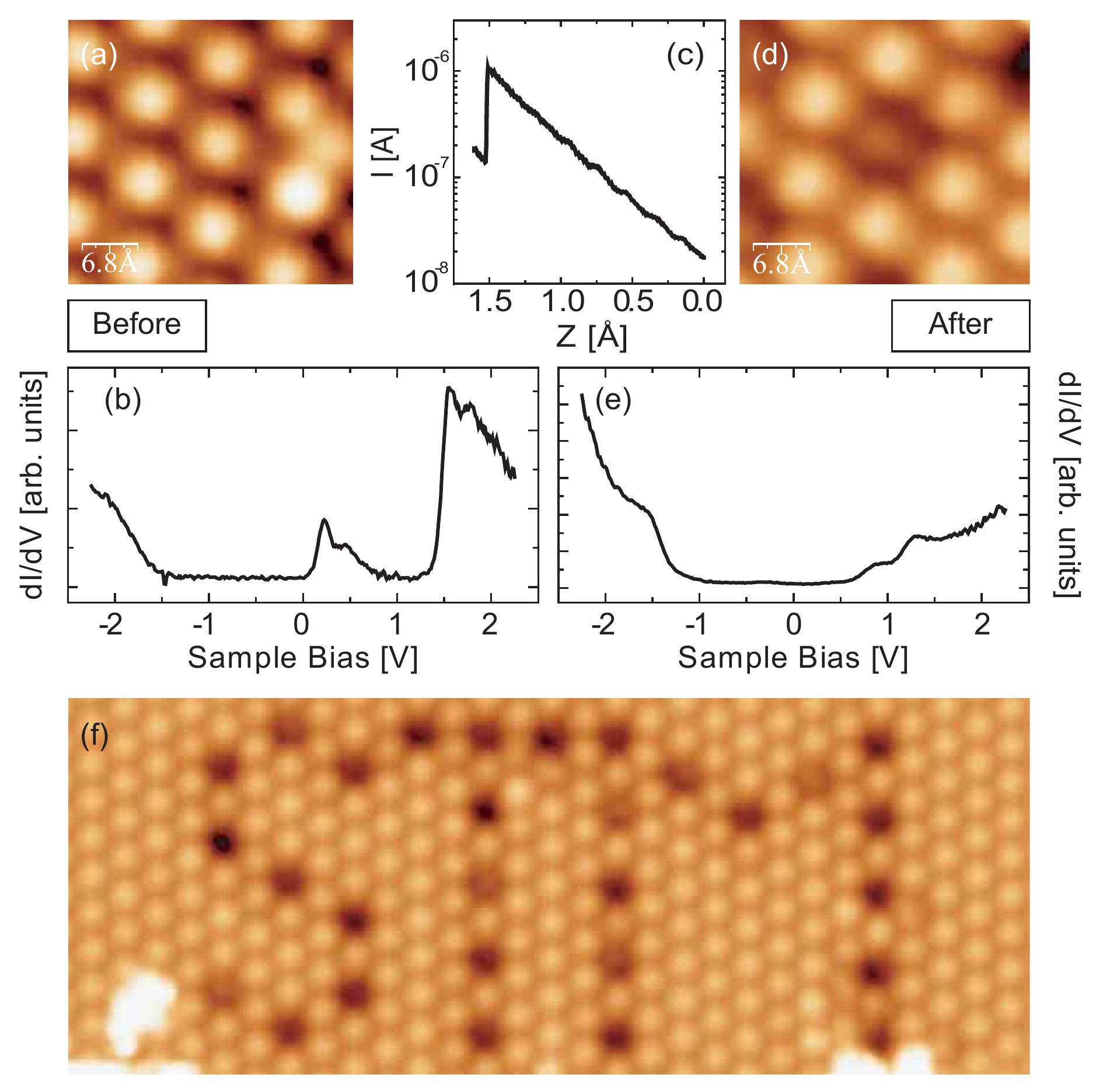}
  \end{flushright}
  \caption{Moleculare decomposition of \CSi\ on a Pb(111) surface.
  (a, b): STM image and \dIdV\ plot of an intact \CSi\ (central molecule).
  (c): Typical I(Z) curve of a tip approach experiment
  showing one decomposition event in the tunnelling regime ($V_s = 2.75\ V$).
  (d, e): STM image and \dIdV\ plot of the molecule from (a, b) after decomposition events
  on the central molecule and on molecules
  at the image border. A height difference of 0.7 \AA\ and the disappearance of the
  characteristic LUMO and LUMO+1 resonances confirms the degradation of the \CSi\ molecule.
  ( (a): $I_t = 0.2\ nA$, $V_s = 0.5\ V$, (d): $I_t = 0.2\ nA$, $V_s = 2.25\ V$,
  (b,e): $R_{junc} = 2.3\ G\Omega$ )
  (f): Image of the letters ``STM"´ created by successive
  decomposition events, to illustrate the large reproducibility ($I_t = 0.2\ nA$, $V_s = 0.5\ V$). }
  \label{Cracking}
\end{figure}

Fig. \ref{Cracking} shows, for the case of Pb(111), the effect of
approaching a molecule at bias voltages above a threshold value of
$V_s = 1.0\ V$. A sharp drop appears in the I(Z) curves before
reaching the contact regime \cite{note2}, denoting the occurrence of
an irreversible change in the junction (Fig. \ref{Cracking} (c)).
Similar results can also be found on Cu(110) for $V_s > 1.2\ V$
\cite{Schulze08} and on Au(111) $V_s > 1.5\ V$. After this current
drop STM images reveal that the molecule is transformed into a lower
feature (Fig. \ref{Cracking} (d)) and the characteristic resonances
of the fullerene icosahedral cavity are absent from its
corresponding STS spectrum (Fig. \ref{Cracking} (e)). Hence, the
discontinuous current drop is a fingerprint of degradation of the
fullerene cage. The precise way in which the \CSi\ decomposes can
not be determined in detail in our measurements. The effect is
observed solely on the molecule selected for the tip approach and
shows a high reproducibility, as depicted in Fig. \ref{Cracking} (f)
for the case of \CSi\ on Pb(111) and as shown in Ref.
\cite{Schulze08}, on Cu(110) \cite{rotation}. A mechanical breaking
of the fullerene cage can be excluded since the process takes places
before a tip-molecule contact is formed (i.e. in the tunneling
regime). We can then conclude that the decomposition is a current
induced thermal process.

\begin{figure}[h]
  \begin{flushright}
    \includegraphics[width = 8cm]{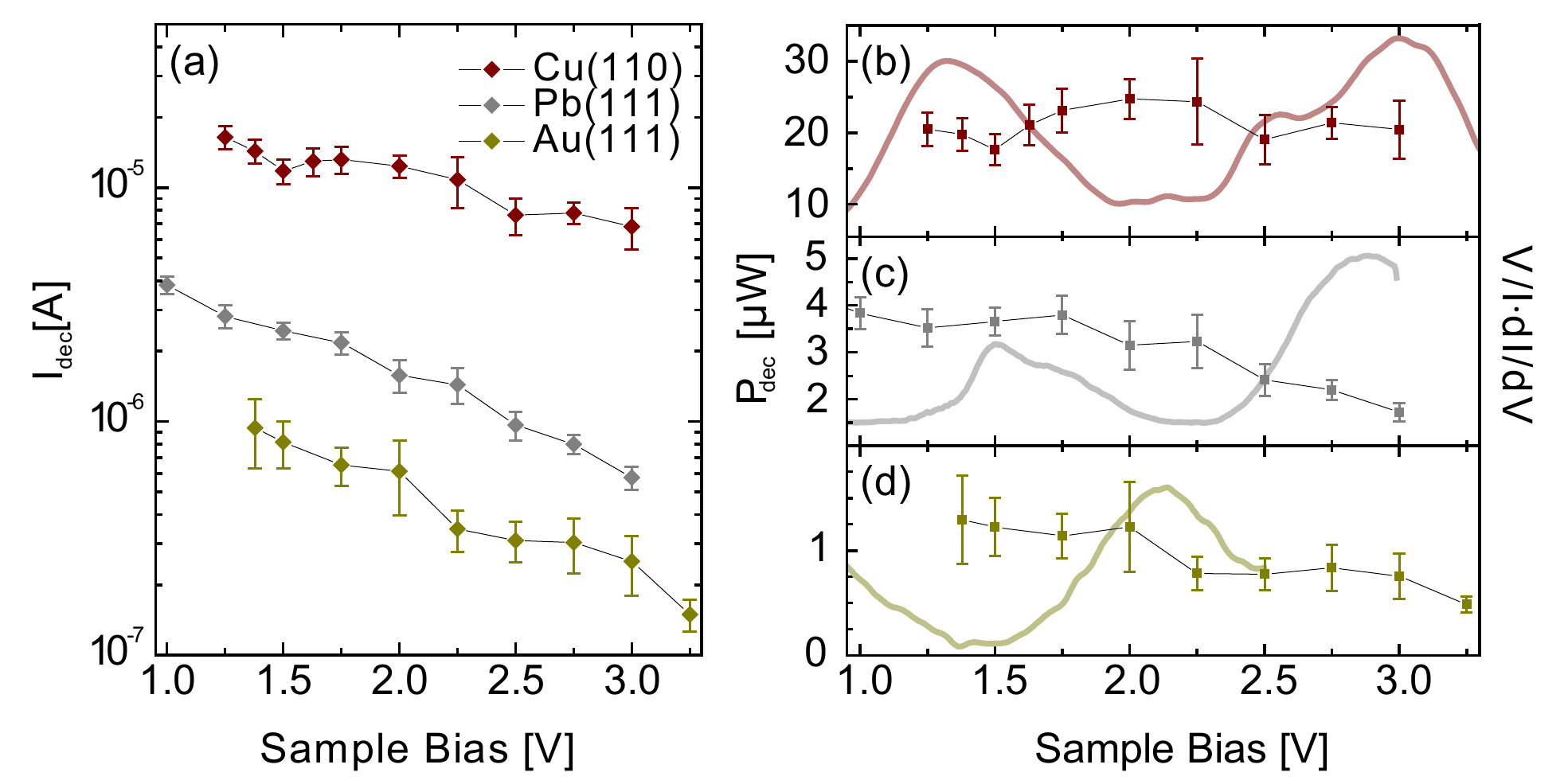}
  \end{flushright}
  \caption{(a): Statistical average of decomposition currents $\rm
  I_{dec}(V_s)$ in the tunnelling regime for the three different surfaces.
  The total numbers of recorded events are: 152 on Cu(110), 150 on Pb(111) and 105 on Au(111).
  (b--d): Statistical average of the decomposition power $\rm P_{dec} (V_s)$ for (b) Cu(110), (c) Pb(111) and
  (d) Au(111). $2\sigma$ error bars are indicated in all figures.
  Normalized dI/dV spectra of \CSi\ on the different substrates
  are added as shaded curves for comparison (Cu: $R_{junc} = 0.9\ G\Omega$,
  Pb: $R_{junc} = 0.7\ G\Omega$, Au: $R_{junc} = 0.3\ G\Omega$). }
  \label{IdecPdec}
\end{figure}

The current \Idec\ at which the \CSi\ decomposition occurs can be
taken from the I(Z) approach curves. \Idec\ turns out to show very
reproducible values  distinctive for every metal substrate and a
clear dependence on the bias voltage used during the tip approach.
Fig. \ref{IdecPdec} shows the statistical average of \Idec\ plotted
vs. \Vs\ for all three surfaces. Adsorbed on Cu(110), \CSi\
molecules can withstand currents of several tens of microamperes,
typically one order of magnitude larger than for \CSi\ on Pb(111) or
Au(111). In general,  \Idec\  decreases  monotonously as the applied
bias is increased. Additionally, some faint steps can be inferred.
This substructure becomes more evident when we plot instead the
applied power necessary to decompose the molecule, $P_{dec}
 = V_s \times I_{dec}$ (Fig. \ref{IdecPdec} (b--d)).

The monotonous decrease of \Idec\ transforms in a fairly flat
behaviour of \Pdec\ with  the applied bias. Fig. \ref{IdecPdec}
(b--d) shows that the average power applied for degradation depends
strongly on the substrate used: \PdecM\ $\sim$ 21 $\mu$W for
Cu(110), \PdecM\ $\sim$ 2.9 $\mu$W for Pb(111), and \PdecM\ $\sim$ 1
$\mu$W for Au(111). Superimposed to these values, the stepped
substructure appears now more pronounced. The structure is
correlated with the alignment of the unoccupied states of the \CSi\
molecule, also shown in Fig. \ref{IdecPdec} (b--d). In general we
find that \Pdec\ decreases whenever a new resonance (here the LUMO+1
and LUMO+2) enters the conduction energy window. Hence, the
molecular resonance structure is reflected in the power needed to
decompose a molecule. Next, we interprete the origin of this
resonant substructure (section 4) and  of the strong substrate
dependence of the applied power for degradation \PdecM\ (section 5).

\section{Molecular heating and cooling mechanisms}
\label{HeatCool}

\begin{figure}[h]
  \begin{flushright}
    \includegraphics[width = 8cm]{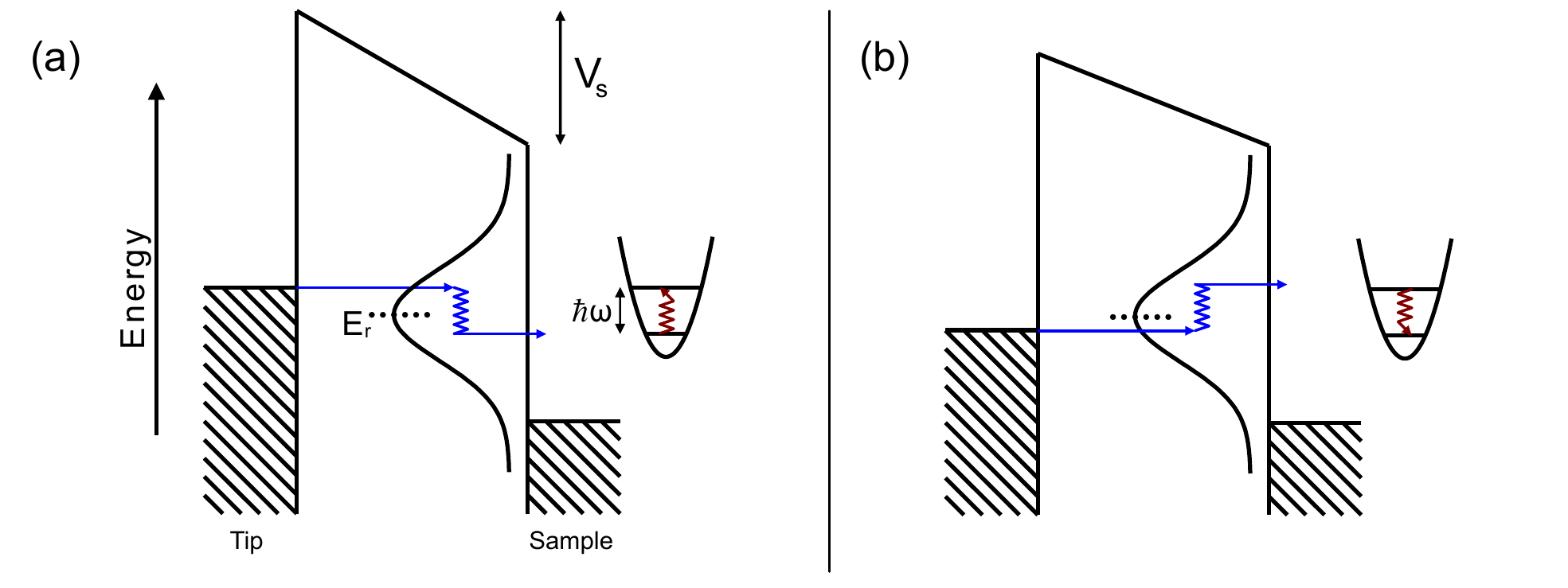}
  \end{flushright}
  \caption{Schematical drawings of vibron interactions with
  tunneling electrons for a molecule inside a tunnel junction: (a) Inelastic
  electron-vibron scattering at a resonance level \Er\ heats the molecule; (b) Vibron
  assisted tunneling is a cooling effect most effective for electron energies close to and below $\rm
  E_r$.}
  \label{SchemesA}
\end{figure}

The origin of the sub-structure in the \Pdec\ vs. \Vs\ plots can be
understood from current theoretical models describing the
electron-induced heating of single molecules during electron
transport \cite{NitzanJPCM07,NitzanPRB07}. Heat inside the molecule
is generated by inelastic  scattering of tunneling electrons with
molecular vibrations (Fig. \ref{SchemesA} (a)). For the large
tunneling rates used in our experiment electron scattering leads to
a non-equilibrium distribution of excited modes, whose internal
energy can be associated with a molecular temperature \Tm. For a
certain set of current and bias values, \Tm\ depends on the balance
between the \textit{heat generated} by the inelastic scattering of
electrons with molecular modes, and \textit{heat dissipated} into
the ``cold" electrode, in our experiment at 5\ K
\cite{PecchiaPRB07,NitzanJPCM07, PecchiaJCM07}.

Recent calculations \cite{PecchiaPRB07} have shown that when a new
resonance level \Er\ enters into the transport energy window (hence,
when $eV_s > E_r$) a steep temperature increase takes place in the
molecule as a consequence of  more vibronic levels being accessible
(Fig. \ref{SchemesA} (a)). Hence, the increase in temperature upon
crossing a resonance level manifests itself as a step-like decrease
of the corresponding decomposition current \Idec\ and power \Pdec,
as shown in the plots of Fig. \ref{IdecPdec}.

In contrast to this, tunnel electrons can also absorb vibrons of a
hot molecule (Fig. \ref{SchemesA} (b)), leading to a reduced heat
generation. When the sample bias lies right below a resonance level
$eV_s \le E_r$) this cooling mechanism can be very effective and,
eventually, causes a lower rise of molecular temperature with \Vs\
\cite{Schulze08}. This behavior translates into plateaus in \Idec\
and increase in \Pdec\ at bias values below the corresponding
resonance energy.

Qualitatively, the combination of both heating and cooling processes
accounts well for the correspondence between the stepped
(oscillating) behavior of the \Idec(\Pdec)  plots and the resonant
structure of the molecule in Fig. \ref{IdecPdec} \cite{Note4}.
However, it cannot explain the striking differences of \PdecM\ found
for the different substrates. Since the mechanisms of
heating/cooling by tunneling electrons depend on the coupling
between tunnel electrons and molecular vibrons they are not expected
to vary much on the different substrates. Hence, the large
differences in \PdecM\ must be related to substrate mediated
mechanisms of cooling the molecule.

\section{Substrate dependence of decomposition power}
\label{substrate}

Further mechanisms of molecular cooling are those in which molecular
hot modes decay by creating quasiparticle excitations in the
substrate (either e-h pairs or phonons). In agreement with our
findings, these excitations  do not depend on the sample bias but
may vary depending on the substrate's electronic and phononic
properties.

\begin{figure}[h]
  \begin{flushright}
    \includegraphics[width = 8cm]{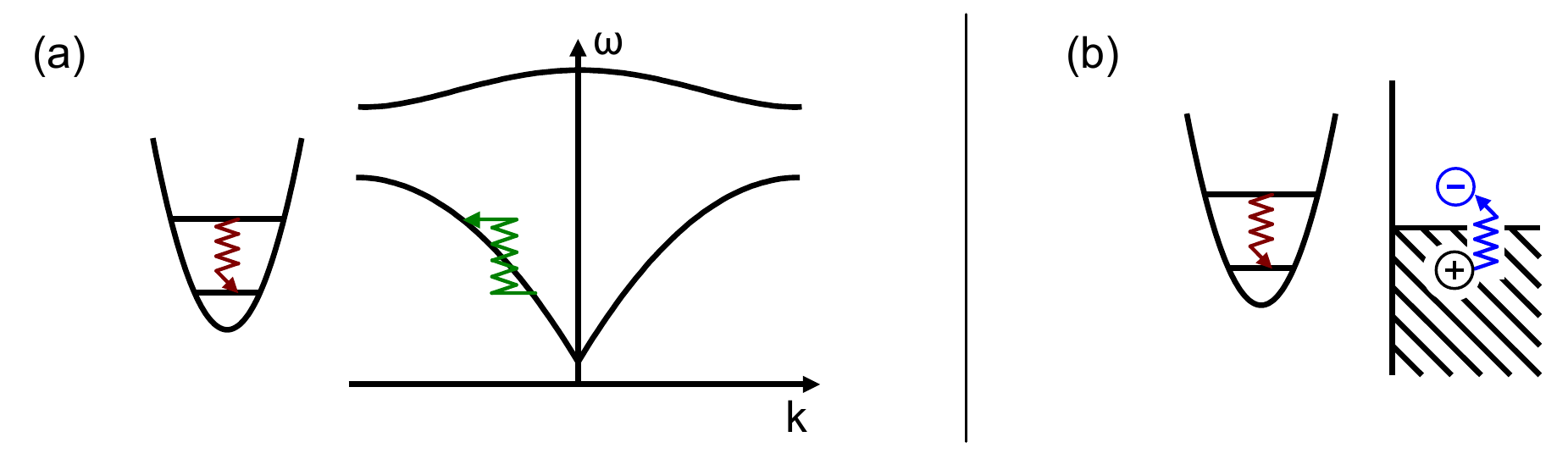}
  \end{flushright}
  \caption{Schematical drawings of mechanisms of molecular vibration
  decay into excitations of the cold metal substrate: (a) decay into the metal phonon band and (b)
  creation of excited metal electrons (electron-hole-pairs) close to the Fermi
  edge are substrate dependent cooling effects.}
  \label{SchemesB}
\end{figure}

Decay of molecular vibrons into substrate phonons  plays a minor
role in cooling the molecule, since this mechanism is limited by the
phonon band width of the substrates (Fig. \ref{SchemesB} (a)). The
Cu(110) phonon band is approximately 42 meV wide and therefore
larger than in Pb(111) (14 meV \cite{Fuhrmann96}) and in Au(111)(16
meV \cite{Wang91}). However, the 174 internal modes of \CSi\ have
energies between 33 meV and 200 meV. Hence, substrate phonons are
expected to be primarily coupled to external molecular vibrations of
the \CSi\ molecule with respect to the surface, which do not
contribute to the thermal decomposition of the fullerene cage.

The most important contribution to the observed substrate dependence
of \PdecM\ is the varying efficiencies of molecular cooling through
electron-hole pair creation (Fig. \ref{SchemesB} (b)) \cite{Gao92}.
Since the final states lie at excitation energies in the order of
the vibrational band width of the molecule, the decay rate scales
with the density of states around \EF. On Cu(110) and Au(111),
surface states provide most of the available states in this energy
region, according to available data of DOS at \EF\ \cite{note3}. It
can therefore be expected, that e-h pair creation is favoured on
Cu(110) with respect to the gold surface, in agreement with our
observations. However, this mechanism fails to explain the
intermediate \PdecM\ value found for degradation of \CSi\ on
Pb(111), since this metal surface has no surface state, and whose
density of bulk states at \EF\ is also the lowest of all three
metals \cite{note3}.

\begin{figure}[h]
  \begin{flushright}
    \includegraphics[width=8cm]{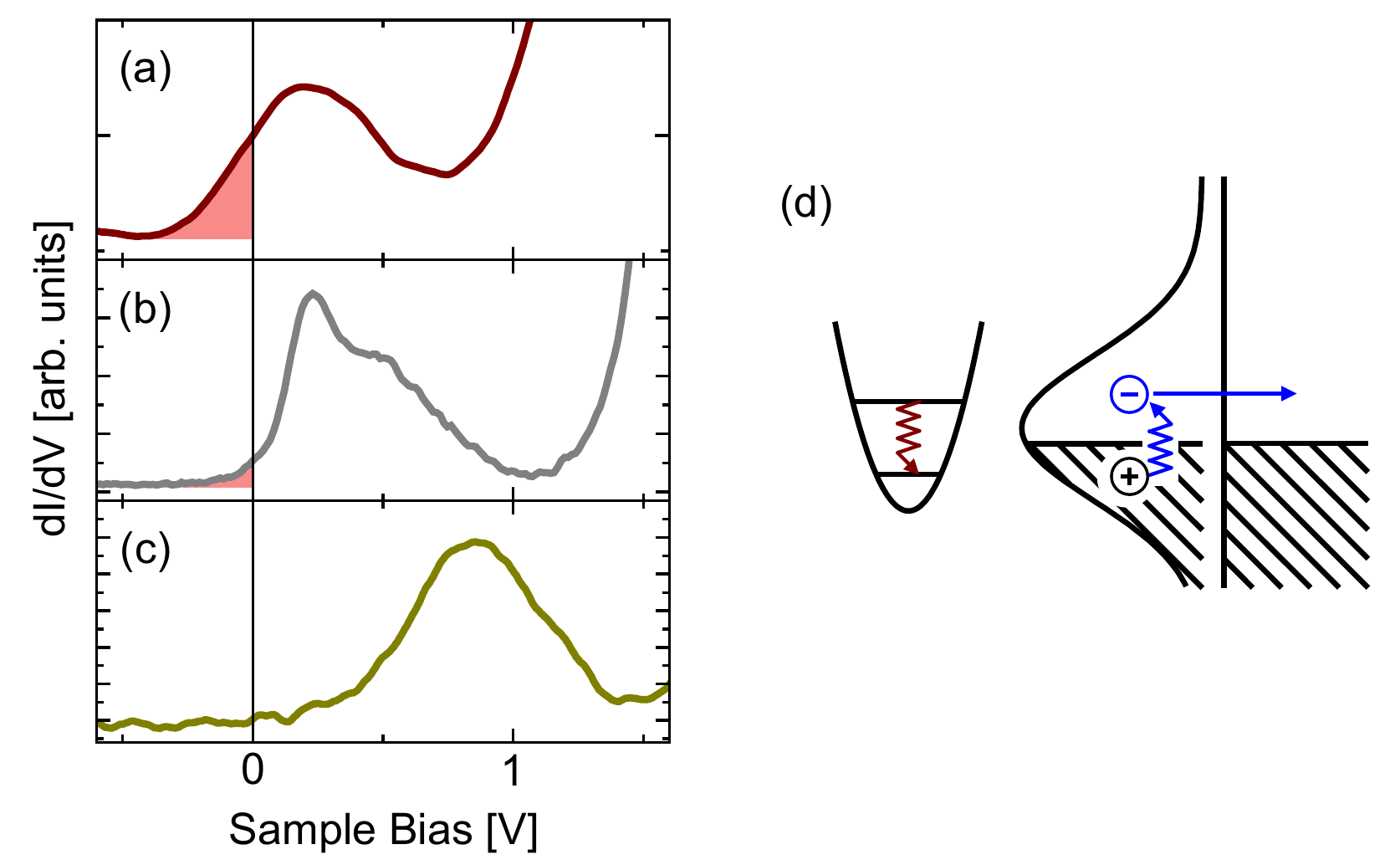}
  \end{flushright}
  \caption{(a--c): Magnification of the \dIdV\ curves of Fig. \ref{ImagedIdV} (d--f) showing the LUMO
  of \CSi\ on (a) Cu(110), (b) Pb(111) and (c) Au(111). The occupied portions of the
  LUMO peaks are marked as shaded areas. From this a decreasing amount
  of electron transfer into the molecule can be expected for
  Cu(110) to Pb(111) to Au(111). (d): Schematical
  drawing of the proposed mechanism of vibron decay by
  electron-hole-pair creation in a partially occupied molecular resonance.
  The large spacial overlap of molecule and metal states at \EF\ increases
  the decay probability.
  The e-h-pairs excited inside the molecule can leave before they recombine.}
  \label{ChargeTransfer}
\end{figure}

To solve this puzzle we note that the absorption of molecular
vibrons through excitation of electron-hole pairs rather  depends on
the density of states around \EF\ \textit{in the \CSi\ molecule
coming from the surface} \cite{Schulze08,AlessioNJP08}. Therefore,
this dissipation channel should not be viewed as an intrinsic
property of the substrate material, but as a consequence of the
molecule-surface interaction. In our case, we could experimentally
determine that the LUMO derived resonances of \CSi\ on different
metals exhibit a different degree of weight at the Fermi energy,
associated  to different amounts of charge transfer from the
surface. In fact, the degree of charge transfer (increasing from
Au(111) over Pb(111) to Cu(110), as shown in Fig.
\ref{ChargeTransfer} (a--c)), follows the trend found for \PdecM\ in
Fig. \ref{IdecPdec}.

This allows us to depict a qualitative picture, in which charge
transfer from the metal causes an increase of the density of
molecular states around \EF. This favours the generation of e-h
pairs at the molecule, which are then reflected back into the metal
where they recombine. The rate of heat dissipation thus increases
with the charge transfer, causing that a larger  power is required
to thermally degrade the single molecule. Ignoring additional
mechanisms and effects like hybridization of metal and molecule
states, which probably could also play an important role here, our
results hint that charging a single molecule in contact with a metal
electrode can help to sustain larger current densities crossing
through a single molecule.

\section{Summary}
\label{summary}

Our experiments reveal that electronic currents in the range of
0.1-20 $\mu$A and powers in the order of 1-30 $\mu$W are sufficient
to generate heat in single \CSi\ molecules leading to thermal
decomposition on metal surfaces. The decomposition power results
from the balance of molecular heating and cooling. While the former
is substrate independent, the latter varies on the different
substrates as a function of charge transfer into the molecule.
Charge transfer assures an effective quenching of molecular vibrons
into electron-hole pairs. In order to increase the current density a
molecular junction can sustain, it is desirable to choose a
substrate where molecular adsorption leads to a partial filling of
molecular states by substrate electrons.

We thank Alessio Gagliardi and Alessandro Pecchia for helpful
discussions. This research was supported by the Deutsche
Forschungsgemeinschaft, through the collaborative projects SPP 1243
and SFB 658.


\section*{References}
\begin{thebibliography}{10}

\bibitem{Galperin07} Galperin M, Ratner M A and Nitzan A 2007 {\it
J. of Phys: Cond. Matt.} {\bf 19} 103201

\bibitem{HuangNL06} Huang Z, Xu B, Chen Y, Di Ventra M and Tao N
2006 {\it Nano Letters}  {\bf 6} 1240

\bibitem{Berndt07a} N\'{e}el N, Kr\"{o}ger J, Limot L, Frederiksen T,
    Brandbyge M and Berndt R 2007 Phys. Rev. Lett. {\bf 98} 065502

\bibitem{HuangNN07} Huang Z, Chen F, D'agosta R, Bennett P A, Di
Ventra M and Tao N 2007 {\it Nature Nanotechnology} \textbf{2} 698

\bibitem{Schulze08} Schulze G, Franke K J, Gagliardi A, Romano G, Lin C, Da Rosa A, Niehaus T A,
Frauenheim Th, Di Carlo A, Pecchia A and Pascual J I 2007 Resonant
electron heating and phonon cooling in single molecule junctions
Phys. Rev. Lett. {\it accepted}

\bibitem{PecchiaPRB07} Pecchia A, Romano G and Di Carlo A 2007 Phys. Rev. B {\bf 75} 035401

\bibitem{FaselPRB99} Fasel R, Agostino R G, Aebi P and Schlapbach L
1999 Phys. Rev. B {\bf 60} 4517

\bibitem{note} Such split structure can be also  obsered on other
substrates but depends on the molecular orientation (e.g. ref.
\cite{FrankePRL07}).

\bibitem{Berndt07c} Schull G and Berndt R 2007 Phys. Rev. Lett. {\bf 99} 226105

\bibitem{PascualJCP02} Pascual J I, Gomez-Herrero J, Sanchez-Portal D,
and Rust H-P 2002 {\it J. Chem. Phys.} {\bf 117} 9531

\bibitem{RogeroJCP02} Rogero C, Pascual J I, Gomez-Herrero J, and Baro A
M 2002 {\it J. Chem. Phys.} {\bf 116} 832

\bibitem{CrommiePRB04}  Lu X, Grobis M, Khoo K H, Louie S G, Crommie M F 2004 Phys. Rev. B {\bf 70} 115418

\bibitem{note2} Below this threshold bias a discontinuity associated to
\CSi\ degradation is still found, but once the tip-molecule contact
has been formed. Below  $V_s = 0.6\ eV$ for Cu(110), $V_s = 0.6\ eV$
for Pb(111), and $V_s = 1.1\ eV$ for  Au(111)) the  \CSi\ molecules
remained intact \cite{Schulze08}.

\bibitem{rotation} It should be noted, that decomposition is not the only possible
process observed during tip approach. Especially rotations of the
molecular orientation are known from other surfaces \cite{Berndt07b}
and can often be found on Au(111). These events exhibit a similar
tunnel current drop behaviour as the decompositions do. They can be
identified by inspection of the \dIdV\ spectra (no serious changes
of the resonances are observed in these cases) and have to be
excluded from the statistics.

\bibitem{NitzanJPCM07} Galperin M, Ratner M A and Nitzan A 2007 {\it J. Phys.:
    Cond. Matt.} {\bf 19} 103201

\bibitem{NitzanPRB07} Galperin M, Nitzan A and Ratner M A 2007 Phys. Rev. B {\bf 75} 155312

\bibitem{PecchiaJCM07} Romano G, Pecchia A and Di Carlo A 2007 {\it J. Phys.: Condens.
Matter} {\bf 19} 215207

\bibitem{Note4} On Pb(111) the LUMO+1 resonance is found
to be split into two peaks at 1.56 V and at 1.78 V. The \Pdec\ step
position at 1.8 V on Pb in Fig. \ref{IdecPdec} is therefore believed
to correspond to the the high energetic part of this resonance.

\bibitem{Fuhrmann96} Fuhrmann D and W\"{o}ll C 1996 {\it Surf. Sci.} {\bf
    368} 20--26

\bibitem{Wang91} Wang X Q 1991 Phys. Rev. Lett. {\bf 67} 1294--1297

\bibitem{Gao92} Gao S, Persson M and Lundqvist B I 1992 {\it Solid
State Comm.} {\bf 84} 271

\bibitem{note3} Density of surface states are approximately
$\rm  0.017 \ (eV \AA^2)^{-1}$  and $\rm 0.0105  \ (eV \AA^2)^{-1}$,
for Cu(110) and Au(111), respectively
\cite{Kevan83,Aebi94,LaShell96}. The bulk DOS   at \EF\ is:
$DOS_{Pb}(E_F) =  0.0166 \ (eV \AA^3)^{-1}$, $DOS_{Au}(E_F) = 0.0173
\ (eV \AA^3)^{-1}$, $DOS_{Cu}(E_F) = 0.025 \ (eV \AA^3)^{-1}$
\cite{Papa86}.

\bibitem{AlessioNJP08} Gagliardi A {\it private communication}

\bibitem{FrankePRL07}  Franke K J, Schulze G, Henningsen N, Fern\'{a}ndez-Torrente I,
Pascual J I, Zarwell S, R\"{u}ck-Braun K, Cobian M, Lorente N 2008
Phys. Rev. Lett. \textbf{100} 036807

\bibitem{Berndt07b} N\'{e}el N, Kr\"{o}ger J, Limot L, Frederiksen T,
    Brandbyge M and Berndt R 2007 Rotation of C60 in a single-molecule
    contact {\it eprint} arXiv:0710.1417

\bibitem{Kevan83} Kevan S 1983 Phys. Rev. B {\bf 28} 4822--4824

\bibitem{Aebi94} Aebi P, Osterwalder J, Fasel R, Naumovi\'{c} D and
    Schlapbach L 1994 {\it Surf. Sci.} {\bf 307--309} 917--921

\bibitem{LaShell96} LaShell S, McDougall B A and Jensen E 1996 Phys. Rev. Lett. {\bf 77}
    3419--3422

\bibitem{Papa86} Papaconstantopoulos D A 1986 {\it Handbook of the
Band Structure of Elemental Solids} (New York: Plenum Press)















\end{thebibliography}
\end{document}